\documentclass[preprintnumbers,amsmath,11pt,amssymb,floatfix,superscriptaddress,noshowpacs,nofootinbib]{revtex4}
\usepackage{graphicx}
\usepackage{epsfig}
\usepackage{bm}
\usepackage{amsfonts}

\input epsf

\begin{document}
\title{\Large A reconstruction of modified holographic Ricci dark energy in $f(R,T)$ gravity}

\author{Antonio Pasqua}
\email{toto.pasqua@gmail.com} \affiliation{Department of Physics,
University of Trieste, Via Valerio, 2 34127 Trieste, Italy.}

\author{Surajit Chattopadhyay}
\email{surajit_2008@yahoo.co.in, surajcha@iucaa.ernet.in}
\affiliation{ Pailan College of Management and Technology, Bengal
Pailan Park, Kolkata-700 104, India.}

\author{Iuliia Khomenko}
\email{ju.khomenko@gmail.com} \affiliation{Heat-and-Power
Engineering Department, National Technical University of Ukraine
Kyiv Politechnical Institute", Kyiv, Ukraine.}

\date{\today}

\begin{abstract}
In this paper, we consider a recently proposed model of Dark Energy (DE) know as Modified Holographic Ricci DE (MHRDE) (which is function of the Hubble parameter and its first derivative with respect to the cosmic time $t$) in the light of the $f\left( R,T \right)$ model of modified gravity, considering the particular model $f\left( R,T \right) = \mu R + \nu T$, with $\mu$ and $\nu$ constants. The equation of state (EoS) parameter $\omega_{\Lambda}$ approaches but never reaches the value -1, implying a quintessence-like behavior of the model. The deceleration parameter $q$ passes from decelerated to accelerated phase at a redshift of $z\approx 0.2$, showing also a small dependence from the values of the parameters considered. Thanks to the statefinder diagnostic analysis, we observed that the $\Lambda$CDM phase for the considered model  is attainable. We observed that the fractional energy densities for DE and DM $\Omega_{\Lambda}$ and $\Omega_m$ have, respectively, an increasing and a decreasing pattern with the evolution of the universe, indicating an evolution from matter to DE dominated universe. Finally, studying the squared speed of the sound $v_s^2$ for our model, we found that is classically stable.
\end{abstract}

\maketitle

\section{Introduction}
The origin of Dark Energy (DE), which is widely believed to be the responsible for the established accelerated expansion of the universe \cite{spe,perlm}, is one of the most serious problems in modern cosmology. An important step toward the comprehension of the nature of DE is to understabd whether it is produced by a cosmological constant $\Lambda$ or it is originated from other sources dynamically changing with time \cite{tsu}. For good reviews on DE see \cite{cople1,pad,bamba1}.\\
In a recent work, Nojiri $\&$ Odintsov \cite{nogiri} discussed the reasons why modified gravity approach is extremely attractive in the applications for late accelerating universe and DE. Another good review on modified gravity was made by Clifton et al. \cite{cli}. Various modified gravity theories have been recently proposed: some of them are $f\left(R\right)$ (with $R$ being the Ricci scalar curvature) \cite{nojiri2,nojiri1}, $f\left( T\right)$ (with $T$ being the trace of stress-energy tensor) \cite{cai,fer,bamba2,bamba3}, Horava-Lifshitz \cite{kir,nis} and Gauss-Bonnet \cite{myr2,ban,nojiri3,li} theories. \\
This work concentrates on $f\left(R, T\right)$ gravity, with $f$ being in this case a function of both $R$ and $T$, manifesting a coupling between matter and geometry. Before going into the details of $f\left(R, T\right)$ gravity, we describe some important features of the $f\left(R\right)$ gravity. The recent motivation for studying $f\left(R\right)$ gravity came from the necessity to explain the apparent late-time accelerating expansion of the Universe. Detailed reviews on $f\left(R\right)$ gravity can be found in \cite{sot1,defelice,sot2,cap}. Thermodynamic aspects of $f\left(R\right)$ gravity have been investigated in the works of \cite{bamba4,akbar}. A generalization of $f\left(R\right)$ modified theories of gravity including in the theory an explicit coupling of an arbitrary function of $R$ with the matter Lagrangian density $L_m$ leads to the motion of massive particles is non-geodesic, and an extra force, orthogonal to the four-velocity, arises \cite{pop}. Harko et al \cite{har} proposed an extension of standard General Relativity, where the gravitational Lagrangian is given by an arbitrary function of $R$ and $T$ and dubbed this model as $f\left(R, T\right)$. The $f\left(R, T\right)$ gravity model depends on a source term, representing the variation of the matter stress-energy tensor with respect to the metric. A general expression for this source term is obtained as a function of $L_m$. Myrzakulov \cite{myr1} recently derived exact solutions for a specific $f\left(R,T\right)$ model which is a linear combination of $R$ and $T$, i.e. $f\left(R, T\right) = \mu R+\nu T$, where $\mu$ and $\nu$ are two constant. Moreover, it was demonstrated that, for some specific values of $\mu$ and $\nu$, the expansion of universe results to be accelerated without introducing any extra dark component.\\
In this work, we consider a recently proposed holographic cosmological model with IR cut-off given by the modified Ricci radius so that $L^{-2}$ is a combination of $H^2$ and $\dot{H}$ (with $H$ and $\dot{H}$ being, respectively, the Hubble parameter and its first derivative with respect to the cosmic time $t$) \cite{chim1,chim2,chim3}. After that, the energy density $\rho_{\Lambda}$ of the MHRDE model becomes:
\begin{eqnarray}
\rho_{\Lambda} = \frac{2}{\gamma - \lambda}\left( \dot{H} + \frac{3\gamma}{2}H^2   \right), \label{1}
\end{eqnarray}
where $\gamma$ and $\lambda$ are free constants.\\
In the limiting case corresponding to  $\left(\gamma =4/3, \lambda =1\right)$ we obtain that $\rho_{\Lambda}$ becomes proportional to the Ricci scalar curvature $R$ for a spatially flat FLRW space-time (corresponding to the curvature parameter $k$ equal to 0). The use of the MHRDE is motivated by the holographic principle because one can relate the DE with an UV cut-off for the vacuum energy with an IR scale such as the one given by $R$. Alternatively, one could proceed in another way by considering $R$ as a new kind of DE, for instance, a geometric DE instead of evoking the holographic principle. Irrespective of the origin of the DE component, it modifies the Friedmann equation leading to a second order differential equation for the scale factor.\\
In this work we are considering MHRDE interacting with pressureless DM with energy density $\rho_m$. Various form of interacting DE models have been constructed in order to fulfil the observational requirements. Plethora of literatures are available where the interacting DE have been discussed. Some examples of interacting DE are presented in \cite{JAMM,wu2,kim2,setmr,wa,karami}.\\
This work aims to reconstruct the MHRDE  under $f\left(R, T\right)$ gravity and it  is organized as follow. In Section 2 we describe the $f\left( R,T   \right)$ model we are considering. In Section 3, we studied the MHRDE model in $f\left( R,T   \right)$ gravity. In Section 4, we have studied the statefinder parameters for the considered model. In Section 5, we discuss the results obtained in the previous Sections. Finally, in Section 6, we write the conclusions of this work.

\section{The $f\left(R,T\right) = \mu R + \nu T$ model}
The metric of a spatially flat homogeneous and isotropic universe in FLRW model is given by:
\begin{eqnarray}
    ds^2=dt^2-a^2\left(t\right)\left[dr^2 +r^2 \left(d\theta ^2 + \sin^2 \theta d\varphi ^2\right) \right], \label{2}
\end{eqnarray}
where $a\left(t\right)$ is a dimensionless scale factor, $t$ is the cosmic time, $r$ is the radial component and $\left(\theta, \varphi\right)$ are the two angular coordinates.\\
We also know that the tetrad orthonormal components $e_i\left( x^{\mu}  \right)$ are related to the metric through the following relation:
\begin{eqnarray}
g_{\mu \nu} = \eta _{ij}e^i_{\mu} e^j_{\nu} \label{metric}
\end{eqnarray}
The Einstein field equations are given by:
\begin{eqnarray}
H^2 &=& \frac{1}{3}\rho,\label{3} \\
\dot{H} &=& -\frac{1}{2}\left(\rho + p  \right), \label{4}
\end{eqnarray}
where $\rho$ and $p$ represent, respectively, the total energy density and the total pressure (choosing $8 \pi G = c = 1$). The conservation equation is given by:
\begin{eqnarray}
\dot{\rho}+3H\left(\rho + p\right)=0. \label{5}
\end{eqnarray}
where:
\begin{eqnarray}
\rho &=& \rho_{\Lambda} + \rho_m,\label{6} \\
p &=& p_{\Lambda}. \label{7}
\end{eqnarray}
We must emphasize that we are considering pressureless DM ($p_m = 0$). Since the components
do not satisfy the conservation equation separately in presence of interaction, we reconstruct
the conservation equation by introducing an interaction term $Q$ which can be expressed in
any of the following forms \cite{shei}: $Q  \propto  H\rho_{\Lambda}$, $Q \propto  H\rho_m$ and $Q \propto  H\left(\rho_m + \rho_{\Lambda}\right)$. In this paper, we consider+ as interaction term the second of the three forms mentioned above. Accordingly, the conservation equation is reconstructed as:
\begin{eqnarray}
\dot{\rho}_{\Lambda}+3H\left(\rho_{\Lambda} + p_{\Lambda}\right) &=& 3H\delta \rho_m, \label{8}  \\
\dot{\rho}_m+3H\rho_m &=& -3H\delta \rho_m. \label{9}
\end{eqnarray}
In Eqs. (\ref{8}) and (\ref{9}), $\delta$ represents the interaction parameter which describes the strength of the interaction between DE and DM and which value is still under debate.\\
One of the most interesting models of $f\left(R, T\right)$ gravity is the so-called $M_{37}$-model, which action $S$ is given by \cite{myr1}:
\begin{eqnarray}
S = \int f\left(  R,T \right)ed^4x + \int L_med^4x, \label{10}
\end{eqnarray}
where $e=det\left( e^i_{\mu} \right)=\sqrt{-g}$ (with $g$ being the determinant of the metric tensor $g_{\mu \nu}$) and $L_m$ is the matter Lagrangian.\\
In this paper, we choose the following expressions for $R$ and $T$:
\begin{eqnarray}
R &=& u + 6\left( \dot{H} + 2H^2  \right), \label{11}\\
T &=& v - 6H^2, \label{12}
\end{eqnarray}
In this paper, we consider the case $u = u\left( a, \dot{a}  \right)$ and $v = v\left( a, \dot{a}  \right)$, where $\dot{a}$ is the derivative of the scale factor with respect to the cosmic time. Moreover, the scale factor $a\left(t\right)$,
the torsion scalar $T$ and the curvature scalar $R$ are considered as independent dynamical variables.
Then, after some algebraic calculations, the action given in Eq. (\ref{10}) can be written as:
\begin{eqnarray}
S_{37} = \int dt L_{37}
\end{eqnarray}
where the Lagrangian $L_{37}$ is given by:
\begin{eqnarray}
L_{37} &=& a^3 \left( f -T f_T -Rf_R +v f_T + uf_R  \right)\\
&&- 6\left(f_R + f_T   \right)a\dot{a}^2  - 6\left(f_{RR} \dot{R} + f_{RT}\dot{T}   \right)a^2\dot{a} -a^3L_m.
\end{eqnarray}
$f_R$, $f_T$, $f_{RR}$ and $f_{RT}$ are, respectively, the first derivative of $f$ respect to $R$, the first derivative of $f$ respect to $T$, the second derivative of $f$ respect to $R$ and the second derivative of $f$ respect to $R$ and $T$.\\
The equations of $f\left(R, T\right)$ gravity are usually more complicated with respect to the equations of General Relativity even if the FRW metric is considered. For this reason, as stated before, we consider the following simple particular model of $f\left(R, T\right)$ gravity:
\begin{eqnarray}
f\left(  R, T \right) = \mu R + \nu T, \label{13}
\end{eqnarray}
with $\mu$ and $\nu$ constants.\\
The equations system of this model of $f\left(R, T\right)$ gravity is given by:
\begin{eqnarray}
\mu D_1 + \nu E_1 + K \left( \mu R + \nu T   \right) &=& -2a^3 \rho, \label{14}\\
\mu A_1 + \nu B_1 + M \left( \mu R + \nu T   \right) &=& 6a^2 p, \label{15}\\
\dot{\rho} + 3H\left( \rho + p  \right) &=& 0,\label{16}
\end{eqnarray}
where:
\begin{eqnarray}
D_1 &=& -6a\dot{a}^2+ a^3u_{\dot{a}}\dot{a}-a^3\left( u-R  \right)  = 6a^2\ddot{a} +a^3\dot{a}u_{\dot{a}}  = a^3\left( 6\frac{\ddot{a}}{a} + \dot{a}u_{\dot{a}} \right),\label{17} \\
E_1 &=&  -6a\dot{a}^2+a^3\dot{a}v_{\dot{a}} -a^3\left( v-T  \right) = -12a\dot{a}^2+a^3\dot{a}v_{\dot{a}} =   a^3\left( -12\frac{\dot{a}^2}{a^2} + \dot{a}v_{\dot{a}}\right), \label{18}\\
K &=& - a^3, \label{19}\\
A_1 &=& 12\dot{a}^2 + 6a\ddot{a} + 3a^2\dot{a}u_{\dot{a}}+a^3u_{\dot{a}} -a^3u_{a},\label{20} \\
B_1 &=& -24\dot{a}^2 -12a\ddot{a} + 3a^2\dot{a}v_{\dot{a}}+a^3v_{\dot{a}} -a^3v_{a},\label{21}\\
M &=& -3a^2.\label{22}
\end{eqnarray}
We get from Eqs. (\ref{14}), (\ref{15}) and (\ref{16}):
\begin{eqnarray}
-6\left( \mu + \nu \right) \frac{\dot{a}^2}{a^2} + \mu \dot{a}u_{\dot{a}} +\nu \dot{a}v_{\dot{a}} -\mu u -\nu v &=& -2\rho, \label{AAA}\\
-2\left( \mu + \nu   \right)\left(\frac{\dot{a}^2}{a^2} + 2\frac{\ddot{a}}{a}   \right)      + \mu \dot{a}u_{\dot{a}} +\nu \dot{a}v_{\dot{a}} -\mu u -\nu v + \frac{\mu}{3}a\left(\dot{u}_{\dot{a}}-u_a\right) + \frac{\nu}{3}a\left(\dot{v}_{\dot{a}}-v_a\right) &=& 2p, \label{BBB}\\
\dot{\rho} + 3H\left( \rho + p  \right) &=& 0.  \label{CCC}
\end{eqnarray}
Then, Eqs. (\ref{AAA}), (\ref{BBB}) and (\ref{CCC}) can be rewritten as follow:
\begin{eqnarray}
3\left( \mu + \nu \right) \frac{\dot{a}^2}{a^2} -\frac{1}{2}\left( \mu \dot{a}u_{\dot{a}} +\nu \dot{a}v_{\dot{a}} -\mu u -\nu v\right) &=& \rho, \label{}\\
\left( \mu + \nu   \right)\left(\frac{\dot{a}^2}{a^2} + 2\frac{\ddot{a}}{a}   \right)     -\frac{1}{2}\left( \mu \dot{a}u_{\dot{a}} +\nu \dot{a}v_{\dot{a}} -\mu u -\nu v \right) - \frac{\mu}{6}a\left(\dot{u}_{\dot{a}}-u_a\right) - \frac{\nu}{6}a\left(\dot{v}_{\dot{a}}-v_a\right) &=& -p, \label{}\\
\dot{\rho} + 3H\left( \rho + p  \right) &=& 0.  \label{}
\end{eqnarray}
or equivalently:
\begin{eqnarray}
3\left( \mu + \nu \right) H^2 -\frac{1}{2}\left( \mu \dot{a}u_{\dot{a}} +\nu \dot{a}v_{\dot{a}} -\mu u -\nu v\right) &=& \rho, \label{AA1}\\
\left( \mu + \nu   \right)\left(2\dot{H} + 3H^2   \right)     -\frac{1}{2}\left( \mu \dot{a}u_{\dot{a}} +\nu \dot{a}v_{\dot{a}} -\mu u -\nu v \right) - \frac{\mu}{6}a\left(\dot{u}_{\dot{a}}-u_a\right) - \frac{\nu}{6}a\left(\dot{v}_{\dot{a}}-v_a\right) &=& -p, \label{AA2}\\
\dot{\rho} + 3H\left( \rho + p  \right) &=& 0.  \label{AA3}
\end{eqnarray}
The above system has 2 equations and 5 unknown functions, i.e. $a$, $\rho$, $p$, $u$ and $v$.\\
We now assume the following expression for $u$ and $v$:
\begin{eqnarray}
u &=& \alpha a^n, \\
v &=& \beta a^m,
\end{eqnarray}
where $m$, $n$, $\alpha$ and $\beta$ are real constants. We also have that $u$ and $v$ can be also expressed as:
\begin{eqnarray}
u &=& \alpha \left( \frac{v}{\beta}  \right)^{\frac{n}{m}}, \\
v &=& \beta \left( \frac{u}{\alpha}  \right)^{\frac{m}{n}}.
\end{eqnarray}
Then, the system made by Eqs. (\ref{AA1}), (\ref{AA2}) and (\ref{AA3}) leads to:
\begin{eqnarray}
3\left( \mu + \nu   \right)H^2 + \frac{1}{2}\left( \mu \alpha a^n + \nu \beta a^m  \right) &=& \rho, \label{23}\\
\left( \mu + \nu   \right) \left( 2\dot{H} +3H^2 \right) + \frac{\mu \alpha \left(n+3\right)}{6}a^n + \frac{\nu \beta \left(m+3\right)}{6}a^m &=& -p, \label{24}\\
\dot{\rho} + 3H\left( \rho + p  \right) &=& 0.    \label{24-1}
\end{eqnarray}
Finally, we have that the EoS parameter $\omega$ for this model is given by the following expression:
\begin{eqnarray}
\omega =\frac{p}{\rho}=-1-\frac{2\left(\mu + \nu \right)\dot{H} -  \frac{\mu}{6}a\left(\dot{u}_{\dot{a}}-u_a\right) - \frac{\nu}{6}a\left(\dot{v}_{\dot{a}}-v_a\right)}  {3\left(\mu + \nu \right)H^2-\frac{1}{2}\left(\mu \dot{a}u_{\dot{a}} +\nu \dot{a}v_{\dot{a}}    -\mu u - \nu v  \right)}.
\end{eqnarray}

\section{INTERACTING MHRDE IN $f\left(R, T\right)$ GRAVITY}
Solving the differential equation for $\rho_m$ given in Eq. (\ref{9}), we obtain the following expression for $\rho_m$:
\begin{eqnarray}
\rho_m = \rho_{m0}a^{-3\left( 1+\delta  \right)}, \label{25}
\end{eqnarray}
where $\rho_{m0}$ represents the present value of $\rho_m$.\\
Using Eqs. (\ref{1}) and (\ref{15}) in the right hand side of the Eq. (\ref{23}), we obtain the following expression of $H^2$ as function of the scale factor $a$:
\begin{eqnarray}
H^2&=&Ca^{3\left[\gamma \left( \mu + \nu -1  \right) - \lambda \left( \mu + \nu \right)   \right]} + \frac{\left(\gamma - \lambda   \right)}{6} \times \nonumber \\
&& \times \left[  \frac{3\beta \nu a^m}{m+3\left[ -\gamma \left(  \mu + \nu -1  \right)  + \lambda \left( \mu + \nu \right) \right]}   +   \right. \nonumber \\
&&\left. \frac{3\alpha \mu a^n}{n+3\left[ -\gamma \left(  \mu + \nu -1  \right)  + \lambda \left( \mu + \nu \right) \right]} +   \right. \nonumber \\
&&\left. \frac{2\rho_{m0} a^{-3\left( 1+\delta \right)}}{1+\delta -\lambda \left( \mu + \nu \right) + \gamma \left(  \mu + \nu -1  \right) }  \right], \label{26}
\end{eqnarray}
where $C$ is a constant of integration.\\
We can now obtain the expressions of the first and the second time derivative of the Hubble parameter $H$, i.e. $\dot{H}$ and $\ddot{H}$, as function of the scale factor $a$ differentiating  Eq. (\ref{26}) with respect to the cosmic time $t$:
\begin{eqnarray}
\dot{H} &=& \frac{3}{2}\left[\gamma \left(\mu + \nu -1  \right)-\lambda \left(\mu +\nu \right)\right]Ca^{3\left[\gamma \left(\mu + \nu -1 \right) -\lambda \left(\mu +\nu \right)\right]} \nonumber \\
&&+\frac{\left(\gamma - \lambda   \right)}{6} \left[  \frac{1}{2}\frac{3m\beta \nu a^m}{m+3\left[ -\gamma \left(  \mu + \nu -1  \right)  + \lambda \left( \mu + \nu \right) \right]}   + \right. \nonumber \\
&& \left. \frac{1}{2}\frac{3n\alpha \mu a^n}{n+3\left[ -\gamma \left(  \mu + \nu -1  \right)  + \lambda \left( \mu + \nu \right) \right]}   \right. \nonumber \\
&&\left. - \frac{3\rho_{m0} \left( 1+\delta  \right)a^{-3\left( 1+\delta \right)}}{1+\delta -\lambda \left( \mu + \nu \right) + \gamma \left(  \mu + \nu -1  \right) }  \right], \label{27}
\end{eqnarray}
\begin{eqnarray}
\ddot{H} &=& \frac{9}{2}H\left[\gamma \left(\mu + \nu -1  \right)-\lambda \left(\mu +\nu \right)\right]^2Ca^{3\left[\gamma \left(\mu + \nu -1 \right) -\lambda \left(\mu +\nu \right)\right]} \nonumber \\
&&+\frac{\left(\gamma - \lambda   \right)H}{6} \left[  \frac{1}{2}\frac{3m^2\beta \nu a^m}{m+3\left[ -\gamma \left(  \mu + \nu -1  \right)  + \lambda \left( \mu + \nu \right) \right]}  \right. \nonumber \\
&& \left. + \frac{1}{2}\frac{3n^2\alpha \mu a^n}{n+3\left[ -\gamma \left(  \mu + \nu -1  \right)  + \lambda \left( \mu + \nu \right) \right]}   \right. \nonumber \\
&&\left. + \frac{9\rho_{m0} \left( 1+\delta \right)^2a^{-3\left( 1+\delta \right)}}{1+\delta -\lambda \left( \mu + \nu \right) + \gamma \left(  \mu + \nu -1  \right) }  \right] . \label{28}
\end{eqnarray}
Using Eqs. (\ref{26}) and (\ref{27}) in Eq. (\ref{1}), we derive the expression of the energy density $\rho_{\Lambda}$ as follow:
\begin{eqnarray}
\rho_{\Lambda}&=&3C\left(\mu +\nu \right)a^{3(\gamma\left(\mu +\nu -1\right)-\lambda\left(\mu +\nu \right))}+\nonumber \\
 &&+\frac{\beta \nu a^m\left( m+3\gamma \right)}{2\left \{ m+3\left[ -\gamma \left(  \mu + \nu -1  \right)  + \lambda \left( \mu + \nu \right) \right]\right \}}    \nonumber \\
    &&+\frac{\alpha \mu a^n\left( n+3\gamma \right)}{2\left \{n+3\left[ -\gamma \left(  \mu + \nu -1  \right)  + \lambda \left( \mu + \nu \right) \right]\right \}}    \nonumber \\
 && + \frac{\rho_{m0} \left( \gamma -1-\delta \right)a^{-3\left( 1+\delta \right)}}{1+\delta -\lambda \left( \mu + \nu \right) + \gamma \left(  \mu + \nu -1  \right) }. \label{29}
\end{eqnarray}
Using the expression of $\rho_{\Lambda}$ given in Eq. (\ref{29}), we get the expression of the  pressure $p_{\Lambda}$ as follow:
\begin{eqnarray}
p_{\Lambda}   &=& - \frac{1}{6}\left[\alpha \mu \left(n+3\right)a^n + \beta \nu  \left(m+3\right)a^m   \right] - \frac{\left(\mu + \nu\right)}{2} \times \nonumber \\
&& \times \left\{  6C\left[1+ \gamma \left(  \mu + \nu -1  \right) -\lambda \left( \mu + \nu  \right)  \right]a^{3\left[ \gamma \left(  \mu + \nu -1 \right) -\lambda \left( \mu + \nu  \right) \right]}   \right. \nonumber \\
&& \left. +\frac{\beta \nu \left(\gamma - \lambda \right) a^m\left( m+3 \right)}{m+3\left[ -\gamma \left(  \mu + \nu -1  \right)  + \lambda \left( \mu + \nu \right) \right]} \right. \nonumber \\
&&\left.  + \frac{\alpha \mu \left(\gamma - \lambda\right) a^n\left( n+3\right)}{n+3\left[ -\gamma \left(  \mu + \nu -1  \right)  + \lambda \left( \mu + \nu \right) \right]}  \right. \nonumber \\
&&\left.-\frac{2\rho_{m0}a^{-3\left( 1+\delta \right)}\delta \left(\gamma - \lambda \right)}{1+\delta -\lambda \left( \mu + \nu \right) + \gamma \left(  \mu + \nu -1  \right)}  \right\}. \label{30}
\end{eqnarray}
Using the expressions of the energy density $\rho_{\Lambda}$ and the pressure $p_{\Lambda}$ of DE given, respectively, in Eqs. (\ref{29}) and (\ref{30}) and the expression of $\rho_m$ given in Eq. (\ref{25}), we get the EoS parameter $\omega_{\Lambda}$ for DE and the total EoS parameter $\omega_{tot}$ as follow:
\begin{eqnarray}
\omega_{\Lambda} &=& \frac{p_{\Lambda}}{\rho_{\Lambda}}, \label{31}\\
\omega_{tot} &=& \frac{p_{\Lambda}}{\rho_{\Lambda}+\rho_m}.\label{32}
\end{eqnarray}
It must be remembered that we are considering pressureless DM, so that $p_m=0$.\\
The deceleration parameter $q$ comes out to be:
\begin{eqnarray}
q= -1 - \frac{a\ddot{a}}{\dot{a}^2} = -1 - \frac{\dot{H}}{H^2}. \label{33}
\end{eqnarray}
where the expressions of $H^2$ and $\dot{H}$ are given, respectively, in Eqs. (\ref{26}) and (\ref{27}).
The deceleration parameter, the Hubble parameter $H$ and the dimensionless energy density parameters $\Omega_{\Lambda}$, $\Omega_m$ and $\Omega_k$ are a set of useful parameters if it is needed to describe cosmological observations.

\section{The statefinder parameters}
In order to have a better comprehension of the properties of the considered DE model, it is useful to compare it with a model independent diagnostics which is able to differentiate between a wide variety of dynamical DE models, including the $\Lambda$CDM model. We will use the diagnostic, also known as statefinder diagnostic, which introduces a pair of parameters $\left \{r, s\right \}$ defined, respectively, as:
\begin{eqnarray}
r &=& 1 + 3\frac{\dot{H}}{H^2}+ \frac{\ddot{H}}{H^3}, \label{34}\\
s&=& -\frac{3H\dot{H}+\ddot{H}}{3H\left( 2\dot{H}+3H^2  \right)}. \label{35}
\end{eqnarray}
Using Eqs. (\ref{26}), (\ref{27}) and (\ref{29}), we get the statefinder parameters as:
\begin{eqnarray}
r &=& 1+ \frac{\rho_1}{\rho_2}, \label{36}\\
s &=& \frac{\zeta _1}{\zeta _2},\label{37}
\end{eqnarray}
with:
\begin{eqnarray}
\rho_1 &=& 18Ca^{3\left[\gamma \left( \mu + \nu -1  \right) - \lambda \left( \mu + \nu \right)   \right]}\left \{\gamma^2 \left( \mu + \nu -1  \right)^2 + \right. \nonumber \\
 &&\left. \lambda \left( \mu + \nu  \right)\left[ -1 +\lambda \left(\mu + \nu \right) \right] - \gamma \left( \mu + \nu -1  \right)\left[ -1 +2\lambda \left(\mu + \nu \right) \right] \right\} \nonumber \\
&&+ \frac{a^m m \left(3 + m\right) \beta \left(\gamma - \lambda \right) \nu}{m + 3 \left[-\gamma \left(-1 + \mu + \nu\right) + \lambda \left(\mu + \nu\right)\right]}  \nonumber \\
&&+\frac{a^n n \left(3 + n\right) \alpha \left(\gamma - \lambda\right) \mu}{n + 3 \left[-\gamma \left( \mu + \nu -1 \right) + \lambda \left(\mu + \nu\right)\right]} \nonumber \\
&&+  \frac{6 a^{-3 \left(1 + \delta \right)} \delta \left(1 + \delta \right) \left(\gamma - \lambda \right) \rho_{m0}}{1 + \delta - \lambda \left( \mu + \nu  \right)  + \gamma \left( \mu + \nu -1 \right)}, \label{38} \\
\rho_2 &=& 4 \left \{  Ca^{3\left[\gamma \left( \mu + \nu -1  \right) - \lambda \left( \mu + \nu \right)   \right]} \right. \nonumber \\
&& \left. + \frac{\left(\gamma - \lambda   \right)}{6} \left[  \frac{3\beta \nu a^m}{m+3\left[ -\gamma \left(  \mu + \nu -1  \right)  + \lambda \left( \mu + \nu \right) \right]}   \right. \right. \nonumber \\
&& \left. \left. + \frac{3\alpha \mu a^n}{n+3\left[ -\gamma \left(  \mu + \nu -1  \right)  + \lambda \left( \mu + \nu \right) \right]} \right. \right.  \nonumber \\
&&\left. \left. + \frac{2\rho_{m0} a^{-3\left( 1+\delta \right)}}{1+\delta -\lambda \left( \mu + \nu \right) + \gamma \left(  \mu + \nu -1  \right) }  \right]   \right\}, \label{39} 
\end{eqnarray}
and
\begin{eqnarray}
\zeta _1 &=&  - 18Ca^{3\left[\gamma \left( \mu + \nu -1  \right) - \lambda \left( \mu + \nu \right)   \right]}\left \{\gamma^2 \left( \mu + \nu -1  \right)^2 + \right. \nonumber \\
 &&\left. \lambda \left( \mu + \nu  \right)\left[ -1 +\lambda \left(\mu + \nu \right) \right] - \gamma \left( \mu + \nu -1  \right)\left[ -1 +2\lambda \left(\mu + \nu \right) \right] \right\} \nonumber \\
&&+ \frac{a^m m \left(3 + m\right) \beta \left(\gamma - \lambda \right) \nu}{m + 3 \left[-\gamma \left(-1 + \mu + \nu\right) + \lambda \left(\mu + \nu\right)\right]}  \nonumber \\
&&+\frac{a^n n \left(3 + n\right) \alpha \left(\gamma - \lambda\right) \mu}{n + 3 \left[-\gamma \left( \mu + \nu -1 \right) + \lambda \left(\mu + \nu\right)\right]} \nonumber \\
&&+  \frac{6 a^{-3 \left(1 + \delta \right)} \delta \left(1 + \delta \right) \left(\gamma - \lambda \right) \rho_{m0}}{1 + \delta - \lambda \left( \mu + \nu  \right)  + \gamma \left( \mu + \nu -1 \right)}, \label{40} \\
\zeta_2 &=& 36 C a^{3\left[\gamma\left(\mu + \nu -1 \right)-\lambda \left( \mu+\nu  \right)\right]}\left[1+\gamma\left(\mu + \nu -1 \right)-\lambda \left( \mu+\nu  \right)\right] \nonumber \\
&&+\frac{6a^m m\beta \nu \left( m+3\right)\left(\gamma - \lambda\right)}{m+3\left[ -\gamma\left(\mu + \nu -1 \right)+\lambda \left( \mu + \nu\right) \right]} \nonumber \\
&&+\frac{6a^n n\alpha \mu \left( n+3\right)\left(\gamma - \lambda\right)}{n+3\left[ -\gamma\left(\mu + \nu -1 \right)+\lambda \left( \mu + \nu\right) \right]} \nonumber \\
&&+\frac{12a^{-3\left(1+\delta\right)} \delta\left(\gamma - \lambda\right)\rho_{m0}}{1+\delta +\gamma\left(\mu + \nu -1 \right)-\lambda \left( \mu + \nu\right) }  . \label{41}
\end{eqnarray}

\begin{figure}[h]
\begin{minipage}{16pc}
\includegraphics[width=16pc]{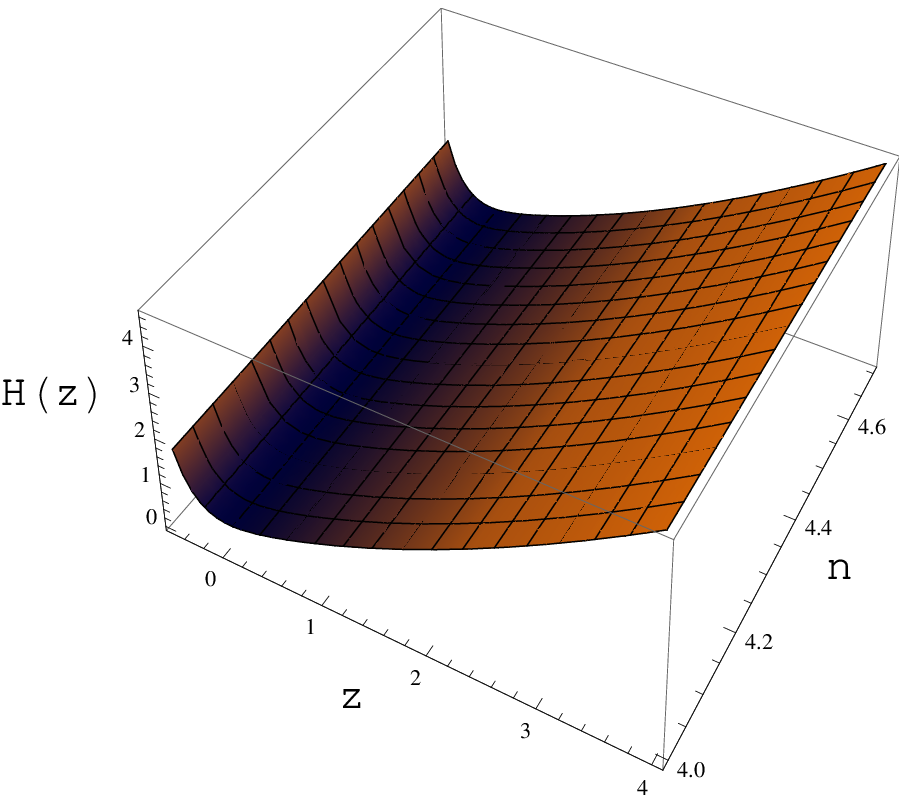}
\caption{\label{label} In this Figure, we plot the Hubble parameter $H$ against redshift $z$
for a range of values of $n$. We have taken
$C=0.01,~\delta=0.05,~\gamma=2,~\nu=0.5,~\mu=0.6,~m=5,~\lambda=0.5,~\beta=0.8,~\alpha=0.5~\textrm{and}~\rho_{m0}=0.23$.}
\end{minipage}\hspace{3pc}%
\begin{minipage}{16pc}
\includegraphics[width=16pc]{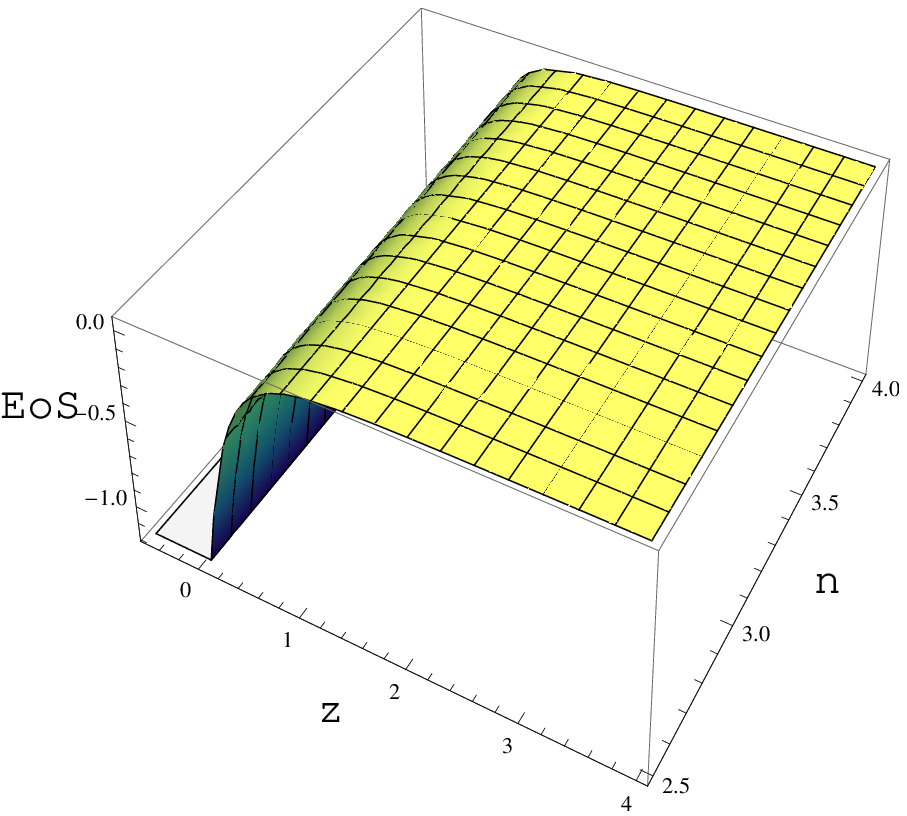}
\caption{\label{label} This Figure plots EoS parameter $\omega_{\Lambda}$ for a
range of values of $n$. We have taken
$C=0.01,~\delta=0.05,~\gamma=2,~\nu=0.5,~\mu=0.6,~m=5,~\lambda=0.5,~\beta=0.8,~\alpha=0.5~\textrm{and}~\rho_{m0}=0.23$.}
\end{minipage}\hspace{3pc}%
\end{figure}

\section{Discussion}
In this Section, we discuss the plots of the physical quantities derived in the previous Section.\\
In Figure 1, we plotted the Hubble parameter $H$ derived from Eq.
(\ref{26}) against the redshift $z$. It is clear that $H$ exhibits a decaying behavior with varying $n$ and $z$ from higher to lower redshifts.\\
In Figure 2, we plotted the EoS parameter $\omega_{\Lambda}$
given in Eq. (\ref{31}) under the considered model of $f\left(R, T\right)$ gravity
against redshift $z$. We observe that $\omega_{\Lambda}$ is
approaching towards $-1$ as the universe evolves from very early
stage (i.e. $z\approx 2$) to the later stages. However, it
crosses the barrier of $-1$. Hence,
we conclude that $\omega_{\Lambda}\geq-1$, which indicates a quintessence-like
behavior of $\omega_{\Lambda}$.\\
\begin{figure}[h]
\begin{minipage}{16pc}
\includegraphics[width=16pc]{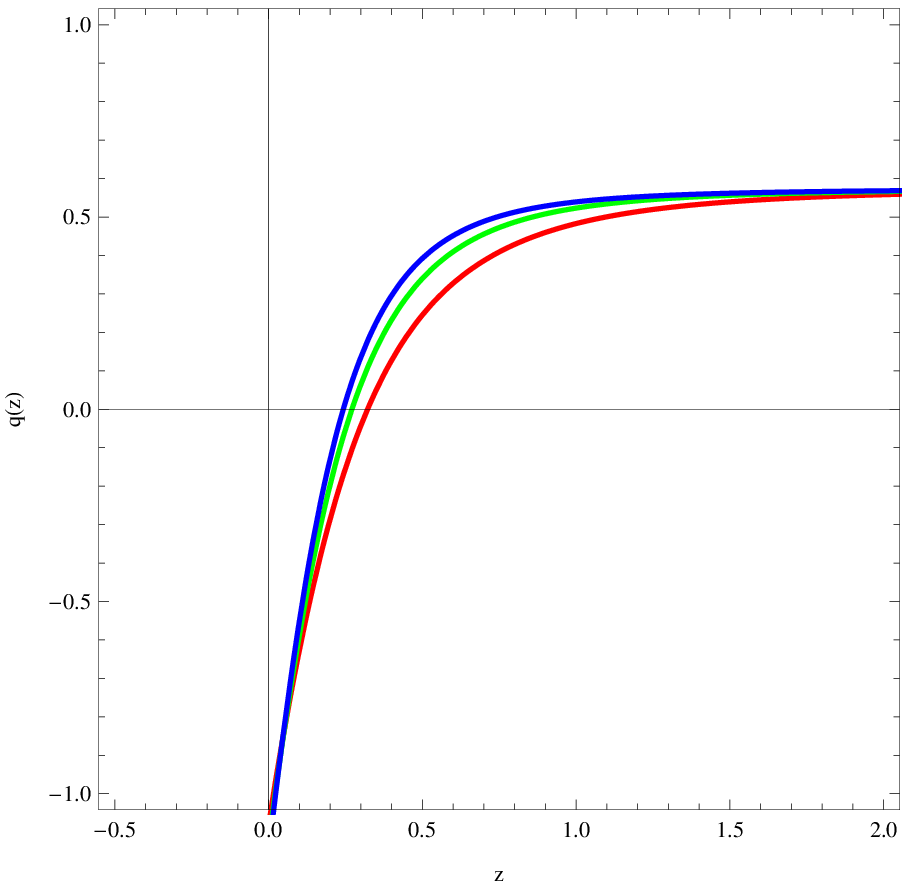}
\caption{\label{label}In this Figure, we plot the deceleration parameter
$q$ given in Eq. (\ref{33}) against redshift $z$ for $n=1.5,~2.5~\textrm{and}~3.5$. We
have taken
$C=0.01,~\delta=0.05,~\gamma=2,~\nu=0.5,~\mu=0.6,~m=5,~\lambda=0.5,~\beta=0.8,~\alpha=0.5~\textrm{and}~\rho_{m0}=0.23$.
The red, green and blue lines correspond to
$n=1.5,~2.5~\textrm{and}~3.5$ respectively.}
\end{minipage}\hspace{3pc}%
\begin{minipage}{16pc}
\includegraphics[width=16pc]{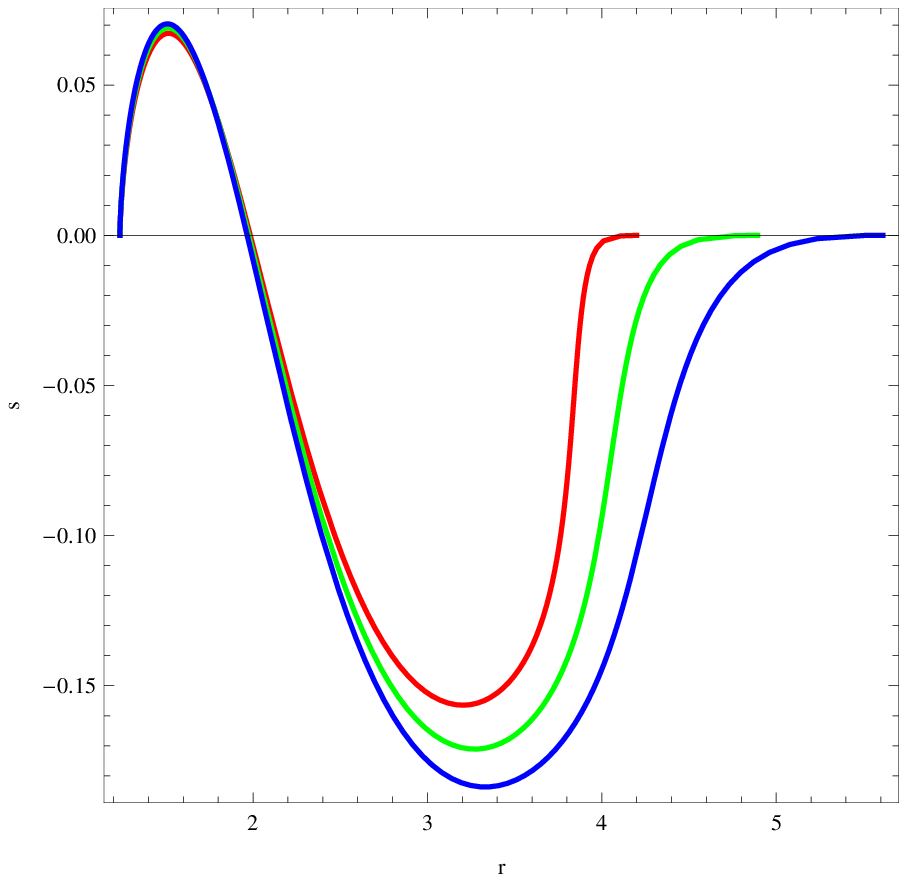}
\caption{\label{label} In this Figure, we plot the $\{r-s\}$ trajectory.
The red, green and blue lines correspond to
$n=1.5,~1.7~\textrm{and}~1.9$ respectively.}
\end{minipage}\hspace{3pc}%
\end{figure}
In Figure 3, we plotted the expression of the deceleration parameter $q$ given
in Eq. (\ref{33}) as a function of the redshift $z$. This plot
shows that $q$ is transiting from positive to negative level at a redshift $z \approx0.2$, which indicates the
transition from decelerated to accelerated phase of the universe.
Furthermore, we observe that the transition gets
delayed as $n$ increases.\\
Sahni et al. \cite{sah} demonstrated that the
statefinder diagnostic can effectively discriminate between different
models of DE. Braneworld, cosmological constant, Chaplygin gas and quintessence models were investigated by Alam et al. \cite{alam} using the
statefinder diagnostics: they observed that the statefinder
pair could differentiate between these different models. An investigation on
statefinder parameters for differentiating between DE and modified
gravity was carried out in \cite{wang}. Statefinder diagnostics for
$f\left(T\right)$ gravity has been studied in Wu $\&$ Yu \cite{wu1}. In
the $\left \{r, s\right \}$ plane, $s > 0$ corresponds to a
quintessence-like model of DE and $s < 0$ corresponds to a phantom-like model of DE. Moreover, an evolution from phantom to quintessence or inverse is given
by crossing of the fixed point $\left(r = 1, s = 0\right)$ in $\left \{r, s \right \}$
plane \cite{wu1}, which corresponds to $\Lambda$CDM scenario. In
Figure 4, we have created the $\left \{r-s\right \}$ trajectories for our model for different values of the
parameter $n$. Clearly, the trajectories attain $\{r=1,~s=0\}$ and, for all
values of $n$, the trajectories converge towards
$\{r=1,~s=0\}$ that pertains to the $\Lambda$CDM phase of the
universe. Hence, we understand that the $\Lambda$CDM phase is
attainable for the interacting MHRDE in $f(R,T)$ gravity.\\
\begin{figure}[h]
\begin{minipage}{16pc}
\includegraphics[width=16pc]{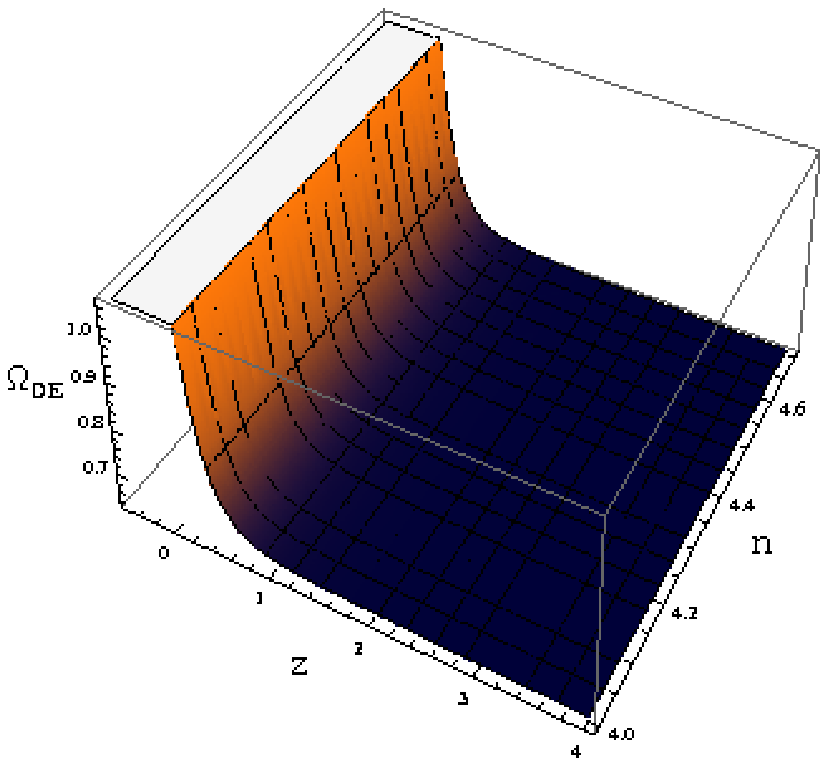}
\caption{\label{label}In this Figure, we plot the fractional DE density
$\Omega_{\Lambda}$ against redshift $z$.}
\end{minipage}\hspace{3pc}%
\begin{minipage}{16pc}
\includegraphics[width=16pc]{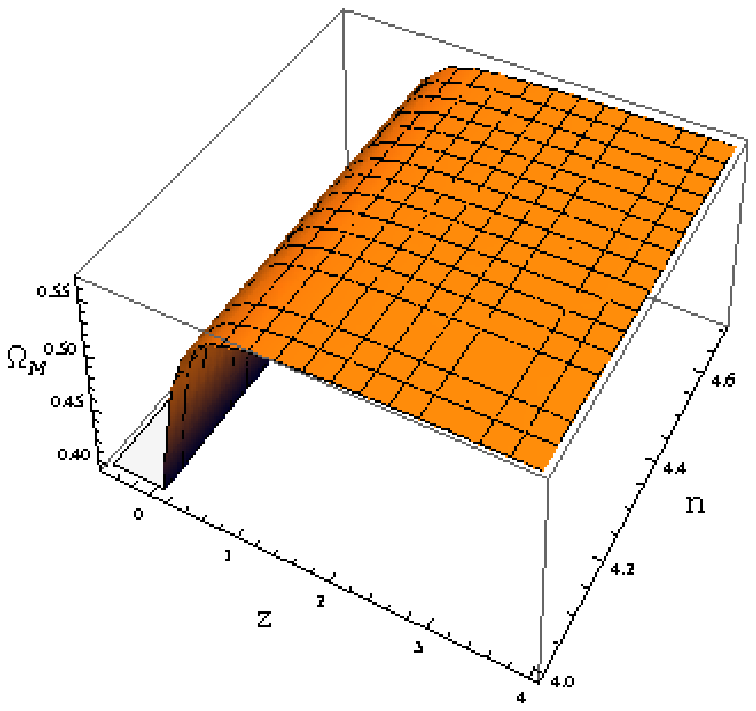}
\caption{\label{label} In this Figure, we plot the fractional dark matter
density $\Omega_{m}$ against redshift $z$.}
\end{minipage}\hspace{3pc}%
\end{figure}

In Figures 5 and 6, we plotted, respectively, the fractional
density of DE $\Omega_{\Lambda} =
\frac{\rho_{\Lambda}}{3\widetilde{H}^2\left(z\right)}$ and the fractional
density of matter $\Omega_{m} = \frac{\rho_m}{3\widetilde{H}^2\left(z\right)}$,
where $\widetilde{H}^2\left(z\right) = \left(\mu + \nu
\right)H^2 + \frac{1}{6}\left[ \alpha \mu \left(1+z \right)^{-n} +
\beta \nu  \left(1 + z \right)^{-m} \right]$, against redshift $z$. Figures 5 and 6 lead us
to conclude that $\Omega_{\Lambda}$ is increasing while $\Omega_{m}$ is decaying with the
evolution of the universe, which indicates the evolution of the
universe from matter dominated to DE dominated universe.\\
\begin{figure}[h]
\includegraphics[width=16pc]{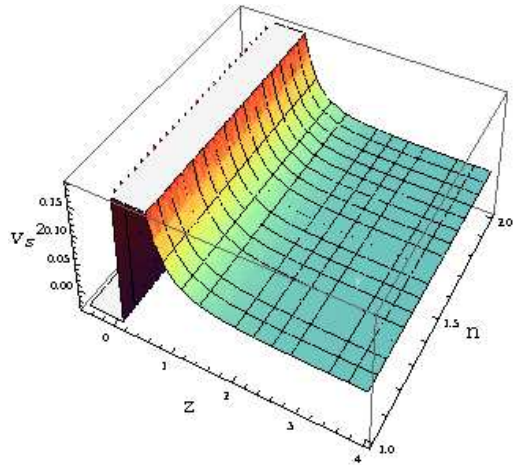}
\caption{\label{label}In this Figure, we plot the squared speed of the
sound $v_{s}^{2}$ against redshift $z$.}
\end{figure}
Finally, in Figure 7, we plotted the squared speed of the sound, defined as
$v_{s}^{2}=\frac{p'}{\rho'}$, where the upper dash indicates
derivative with respect to $\ln a$. In Figure 7, we
plot $v_{s}^{2}$ for our model as a function of $z$ and with varying $n$. We
observe that, throughout the evolution of the universe, the
$v_{s}^{2}$ is staying at positive level: this indicates that the
MHRDE in $f(R,T)$ gravity is classically stable.
\\

\section{Concluding remarks}
In this work, we considered a recently proposed model of energy density of DE known as Modified Holographic Ricci DE (MHRDE) interacting with pressureless DM in the framework of the $f\left(R,T\right)$ modified gravity theory for the special model given by $f\left(R, T\right) = \mu R + \nu T$, where $\mu$ and $\nu$ represents two constants.\\
We derived the expressions and studied the behavior of some important physical quantities which gave useful hints about the model studied. The Hubble parameter $H$ is found to have a decaying behavior going from lower to higher values of the redshift $z$ for all values of the parameter $n$ considered. The EoS parameter $\omega_{\Lambda}$ always assumes values greater than -1, which tell us that the EoS parameter has a quintessence-like behavior for our model. The deceleration parameter $q$ passes from decelerated to accelerated phases of the universe at a redshift $z\approx 0.2$. Moreover, we observe that the value of the transition redshift is a bit delayed as the values of the parameter $n$ increases. Using the statefinder diagnostic, we could find that, for all values of the parameter $n$, the $\Lambda$CDM
 results to be attainable for the interacting MHRDE model. The plots of the fractionale energy densities of DE and DM $\Omega_{\Lambda}$ and $\Omega_m$ show us that the evolution of the universe goes from matter dominated to DE dominated. Finally, we observed that the squared speed of the sounds $v_s^2$ stays at a positive level throughout the evolution of the universe, indicationg that the interacting MHRDE model in $f\left( R,T \right)$ gravity is classically stabel.

\section{Acknowledgement}
The second author (SC) sincerely acknowledges the facilities
provided to him by the Inter-University Centre for Astronomy and
Astrophysics (IUCAA), Pune, India, during his visit in January,
2013 under the Visitor Associateship Programme. Financial support
from DST, Govt. of India under project SR/FTP/PS-167/2011 is duly
acknowledged by SC.

\end{document}